\documentclass{article}

\usepackage{arxiv}

\usepackage[utf8]{inputenc} 
\usepackage[T1]{fontenc}    
\usepackage[pagebackref=true,breaklinks=true,colorlinks,bookmarks=false]{hyperref}
\usepackage{url}            
\usepackage{booktabs}       
\usepackage{amsfonts}       
\usepackage{nicefrac}       
\usepackage{microtype}      

\usepackage{graphicx}
\graphicspath{ {./images/} }

\usepackage{times}
\usepackage{epsfig}
\usepackage{graphicx}
\usepackage{amsmath}
\usepackage{amssymb}

\usepackage{cleveref}
\usepackage{algorithm}
\usepackage{algorithmic}
\usepackage{multirow}
\usepackage[caption=false]{subfig}
\usepackage{adjustbox}
\usepackage{caption}
\usepackage{xcolor}

\usepackage{xspace}

\usepackage{siunitx}
\usepackage{pgf,tikz,pgfplots}
\pgfplotsset{compat=1.15}
\usepackage{mathrsfs}
\usepackage{wrapfig}
\usepackage{lscape}
\usepackage{rotating}
\usepackage{epstopdf}
\usepackage{cleveref}
\usepackage{import}
\usepackage{pdfpages}
\usepackage{booktabs}
\usepackage{adjustbox}

\usepackage{tikz}
\usepackage{xcolor}

\DeclareRobustCommand{\circled}[1]{\tikz[baseline=(myanchor.base)] \node[circle,fill=.,inner sep=1pt] (myanchor) {\color{-.}\bfseries\footnotesize #1};}

\newcommand\inputpgf[2]{{
\let\pgfimageWithoutPath\includegraphics
\renewcommand{\pgfimage}[2][]{\pgfimageWithoutPath[##1]{#1/##2}}
\input{#1/#2}
}}
\usetikzlibrary{arrows}

\renewcommand{\c}[1]{\text{\texttt{#1}}}

\makeatletter
\DeclareRobustCommand\onedot{\futurelet\@let@token\@onedot}
\def\@onedot{\ifx\@let@token.\else.\null\fi\xspace}

\makeatother

\title{Hey Alexa what did I just type?\\Decoding smartphone sounds with a voice assistant}

\author{
 Almos Zarandy, Ilia Shumailov and Ross Anderson \\
  Computer Laboratory\\
  University of Cambridge\\
  \texttt{name.surname@cl.cam.ac.uk} \\
}

\begin{document}
\maketitle
\begin{abstract}
Voice assistants are now ubiquitous and listen in on our everyday lives. Ever since they became commercially available, privacy advocates worried that the data they collect can be abused: might private conversations be extracted by third parties? In this paper we show that privacy threats go beyond spoken conversations and include sensitive data typed on nearby smartphones. Using two different smartphones and a tablet we demonstrate that the attacker can extract PIN codes and text messages from recordings collected by a voice assistant located up to half a meter away. This shows that remote keyboard-inference attacks are not limited to physical keyboards but extend to virtual keyboards too. As our homes become full of always-on microphones, we need to work through the implications.
\end{abstract}

\section{Introduction}

Voice assistants like the Amazon Echo and Google Home are now common household devices. They
come with multiple microphones that are always on. They appear to wake only when their
activation command is spoken, but they process all the audio they hear, in order to perform wake-up or keyword detection. For example, Amazon Alexa responds to `Alexa`, whereas Google home responds to `Hey Google`. To improve user experience it is imperative to minimise both false
positive and false negative activations. That is turn means that the wake-word detector sends a lot of audio data to the server. 

This raises privacy concerns as the microphones often pick up confidential or sensitive personal information. In fact, sometimes up to a minute of audio is uploaded to the server without any keywords present~\cite{dubois2020speakers}. The vendors, Amazon and Google, have business models based on collecting and
processing user data~\cite{BBC2020AmazonHB}. 

There have been cases of the devices accidentally
forwarding sound to the developers or another person's device~\cite{Guardian2018AmazonAR}.
Furthermore, the devices often run third-party apps, many of which control other household smart
devices. As the user interface is purely voice-based, third-party apps need to define a voice-only
interaction. 

The smart speaker framework processes and parses voice input, which can be passed to the apps. So app developers don't have to design their own voice recognition tools, and for most purposes
don't need to access raw audio input. Given data protection concerns, there is a debate about
whether apps should be able to access raw audio at all. In the smartphone world, apps that need
microphone access must get permission from the user, though many users will grant any
permission an apps asks for. In the smart speaker world this model would not work, as the user might assume that
every app has a legitimate need for microphone access. There is no general consensus yet, but Amazon
and Google have devised ways of minimising data exposure to third parties. 

Amazon Alexa does not allow third-party apps (called Alexa Skills) to access sound input, or even the
conversation transcript. It parses the voice inputs and provides a high-level API to developers for
designing conversations. Google has a similar system called Home Actions, though it does share the
transcript with the third party. Both Amazon and Google collect and store all voice interactions
(including audio), which users can review on their websites. Employees and contractors around the
world are given access to voice recordings for labelling~\cite{Bloomberg2019AmazonWA}. 

\begin{figure*}
    \centering
    \includegraphics[width=\linewidth]{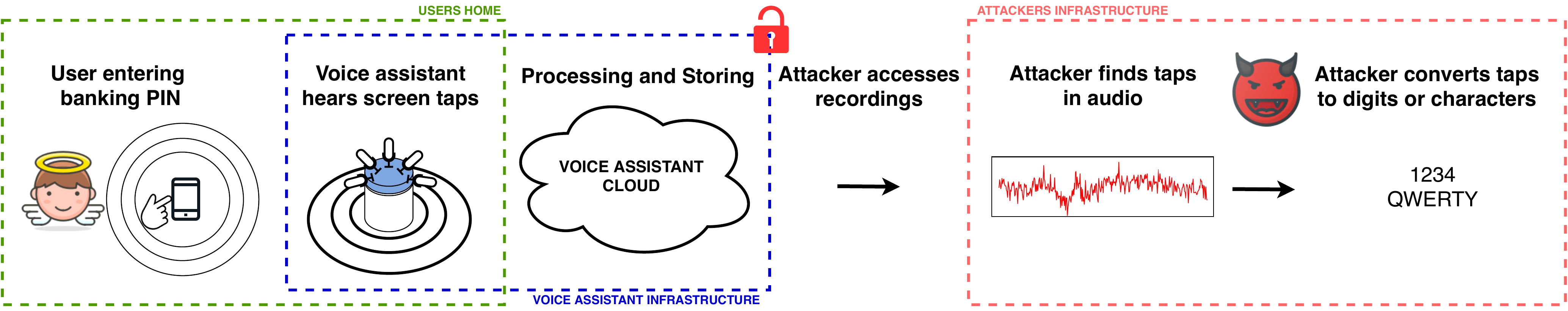}
    \caption{Overall attack flow. User is sitting next to their voice assistant (VA).  \circled{1}~User enters their banking PIN, when VA is active. \circled{2}~VA sends recording of user typing to the cloud where it is processed and stored. \circled{3}~Attacker finds a way to access the audio. \circled{4}~Attacker finds tap sounds in audio and extracts them. \circled{5}~Finally, attacker converts extracted taps into a PIN code. }
    \label{fig:flow_diagram}
\end{figure*}

There has been some discussion of smart speakers overhearing conversations. A survey by
Lau et al.~\cite{Lau2018AlexaAY} finds privacy concerns were a major factor deterring adoption
of smart speakers. Proponents believe that their conversation is not interesting enough for anyone
listening, despite evidence to the contrary. 

In this paper, we study data that smart speakers capture alongside human conversations --
data that users do not expect to be heard. 

Physical keyboards emit sound on key presses. It is well known that captured recording of keystrokes
can be used to reconstruct the text typed on a keyboard. Recent research~\cite{Shumailov2019HearingYT,
Narain2014SingleStrokeLK, Kim2020TapSnoopLT} shows that acoustic side channels can also be exploited with virtual
keyboards such as phone touchscreens, which despite not having moving parts still generate sound. 
The attack is based on the fact that microphones located close to the screen can hear screen
vibrations and use them successfully reconstruct the tap location. Such attacks used to assume
that the adversary could get access to the microphones in the device. We take the attack one step further and relax this assumption.  

In this work we show that attacks on virtual keyboards do not necessarily need to assume access to
the device, and can actually be performed with external microphones. For example, we show how keytaps performed on a smartphone can be reconstructed by nearby smart speakers. The full attack flow is shown in Figure~\ref{fig:flow_diagram}.

Our contributions in the paper are as follows:

\begin{itemize}
    \item we show that it is possible to hear virtual keyboard taps with microphones located on nearby devices;
    \item we show that it is possible to identify PINs and text typed into a smartphone. Given just 10 guesses, 5-digit PINs can be found up to \SI{15}{\percent} of the time, and text can be reconstructed with \SI{50}{\percent} accuracy;
    \item we show that the performance of our early attack is half that achieved with on-device microphones;
    \item we discuss potential mitigation techniques.
\end{itemize}

\section{Threat Model} \label{sec:threat_model}


We assume that an attacker has access to microphones on a 
smart speaker near a target and wants to infer PINs and passwords entered on their touchscreen. There are many ways an attacker might gain access, for example, 
\begin{itemize}
    \item if an attacker installs a malicious skill on a smart speaker and the API allows direct microphone access (not the case for Echo and Google Home unless they can jailbreak the device);
    \item if an attacker gets a job on speech processing at the company producing the device;
    \item if an attacker makes a call to the smart speaker;
    \item if an attacker tampers with the smart speaker;
    \item if an attacker places one or more hidden microphones physically near to the victim;    
    \item if an attacker gets access to raw audio logs;
    \item if they get a job at the audio processing company.
\end{itemize}

We make some further assumptions. First, the phone or tablet is held close to the microphones so that the attacker can guess
its approximate position and orientation, but it does not have to be absolutely still. Second, the make and
model of the device are known and the attacker has access to an identical device which to
obtain training data. An adversary might learn this information by monitoring the victim in various ways. The attacker may give a smart speaker to the victim as a present, and use social engineering to
make them enter the desired information near its microphones. 

\section{Related Work}

\subsection{Acoustic side-channel attacks}

Side-channel attacks against keyboards range from observing the user, through the keyboard itself,
to attacks on the host computer and the network; they are summarised in a systematisation-of-knowledge paper by Monaco~\cite{Monaco2018SoKKS}. 

The first
acoustic attacks were described by Asonov and Agrawal~\cite{Asonov2004KeyboardAE}. They collected
labelled keystroke data from a keyboard, extracted Fourier-transform features, and trained a neural
network to extract keys. 
Zhuang et al.~\cite{Zhuang2005KeyboardAE} improved their attack by using
cepstrum features and linear classification. They also got rid of the need for a labelled
dataset, and used a language model for unsupervised training on data captured from the victim. Their
attacks required access to a microphone near the physical keyboard. 

Cai and Chen were amongst the first to attack virtual keyboards utilising motion sensors~\cite{Cai2011TouchLoggerIK}). Simon and Anderson used the microphone to find when tap events are happening on the screen~\cite{simonanderson2013pinskimmer}. Narain et al.~\cite{Narain2014SingleStrokeLK} used a combination of microphones and gyroscopes to infer key presses on a malicious virtual keyboard. Shumailov et al.~\cite{Shumailov2019HearingYT} presented an attack that only uses microphones. They used a combination of Time Difference of Arrival (TDoA) and raw input classification. Author note that acoustic-only attacks are possible because touchscreens vibrate under finger pressure, which microphones in mechanical contact with the screen can hear. Sound waves propagating through the screen bounce off edges, creating unique patterns that can be reconstructed. 

Cheng et al. showed that swiping passwords can also be cracked with acoustics~\cite{cheng2019sonarsnoop}. They turn the smartphone into an active radar -- an inaudible carrier wave is generated using the speakers and is picked up with the microphones. As a finger moves across a screen it causes Doppler effect, which can be used to reconstruct the swipe direction. 

These attackers relied on access to the microphone data through a malicious app, active phone call or sensor logs. In this work we take a step further and lift this assumption. We show that virtual keyboard inference attacks can be performed with outside microphones, which are a lot more realistic. Indeed, it is common to find voice assistants in a home, and increasingly people's homes are full of always-on microphones. 

\subsection{Voice assistants and mis-activations}

Recent work has highlighted privacy threats from always-on voice assistants. They have been extensively researched both in academia~\cite{abdi2019more,malkin2019privacy,zeng2017end,edu2019smart,bugeja2016privacy,ammari2019music,schonherr2020unacceptable} and by the media~\cite{alexa_privacy_guardian}. Dubois et al.~\cite{dubois2020speakers} analysed misactivations with newscasts and TV shows and found 0.95 per hour, lasting for at least 10s. 
The bottom line is that voice assistants end up capturing a lot more information than intended -- information that can later be mined for market intelligence. In this paper we show that PIN codes and texts entered on a smartphone can often be extracted from such data. 

There have been some suggested defences, but they have not been not widely adopted~\cite{cheng2018towardsacousticjamming}.

\begin{figure*}[h]
	\centerline{%
		\includegraphics[width=0.3\textwidth]{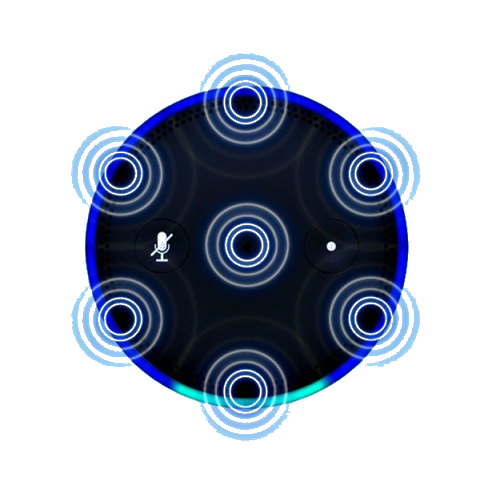}
		\includegraphics[width=0.3\textwidth]{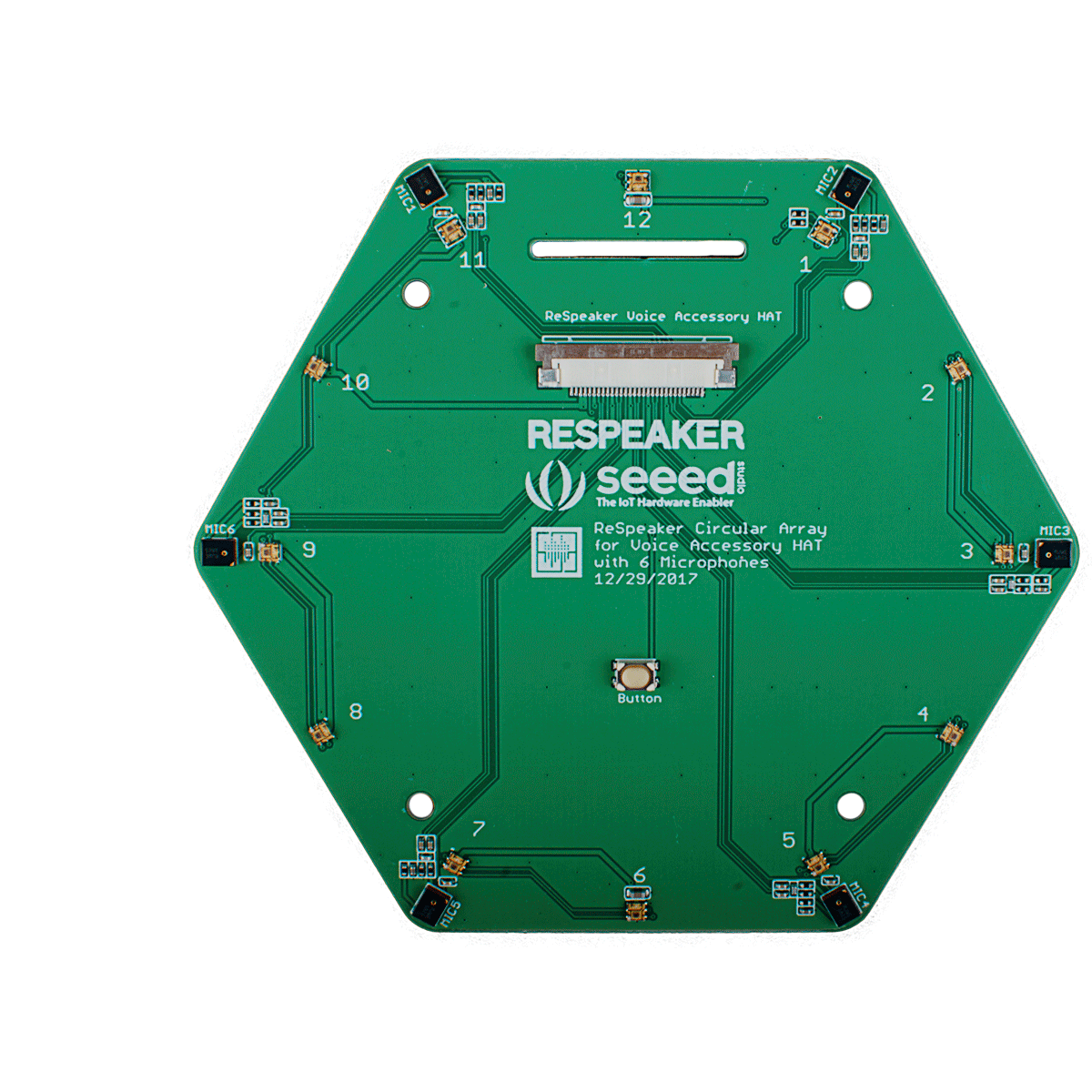}
	}
	\caption{Microphone layout on an Amazon Echo and the ReSpeaker 6mic circular array}
	\label{fig:mic_array}
\end{figure*}

\begin{figure*}[h]
	\centering
	\scalebox{0.6}{%
		\input{figs-gen/waveform-1585646290286-98574-360-.pgf}
		\includegraphics{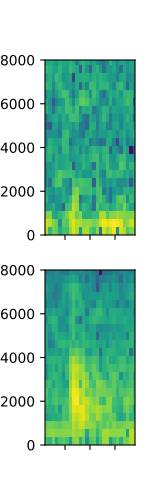}
	}
	\caption{Waveform and spectrogram of a tap on a smartphone screen recorded by the smartphone itself (bottom) and the nearby microphone array (top)}
	 \label{fig:tap_wave_spec}
\end{figure*}

\section{Background}
\label{sec:background}

The Amazon Echo has seven MEMS (Micro Electrical-Mechanical System) microphones on its top
plane, one in the centre and six in a circle on the perimeter, so they can
determine the direction of a sound source. By contrast, the Google Home only has two microphones.

As neither the Echo nor the Home give access to raw audio, our experiments used a ReSpeaker circular
array with 6 microphones. This is an extension to the Raspberry Pi designed to allow running
Alexa on the Pi, but also provide raw audio access with a sampling frequency of \SI{48}{\kilo\hertz}.
The microphones are spaced equally on a circle of radius \SI{4.63}{\centi\metre}.
\Cref{fig:mic_array} shows the microphones on the vertices of the hexagon. The
setup is similar to Alexa, except for its centre microphone. MEMS microphones are small, cheap, extremely sensitive and easy to hide, making them useful for eavesdropping devices. Although the setup is not directly identical to modern voice assistants, it is based on same quality of technology and it is sufficiently similar to explore possible attack vectors.

To localise sound sources in space, a co-ordinate system has to be defined. Our co-ordinate system is centred on hexagon with two of the microphones lying on
the \(x\)-axis. The \(z\)-axis is orthogonal to the plane of the array. 
We use polar co-ordinates: the azimuth \(\theta\) is the angle of the
projection of the point to the plane from the \(x\)-axis, the elevation \(\varphi\) is the angle from
the \(z\)-axis, and the range \(r\) is the distance from the origin. 
In our experiments, the victim device was towards the negative direction of the \(y\)-axis, so the
bottom two microphones were closest to it. 

With a sampling frequency of \SI{48}{\kilo\hertz}, it takes about six samples for sound to travel from a microphone to an adjacent one on the array. 


Taps in the audio recording, both on the victim device and the external array, can be recognised by
a short \SIrange{1}{2}{\milli\second} (50-100 sample) spike with frequencies between
\SIrange{1000}{5500}{\hertz},
followed by a longer burst of frequencies in and around \SI{500}{\hertz}. As these latter frequencies
are common in the background noise of a standard room, it is the initial spike that best
distinguishes taps from noise. 

For example, \Cref{fig:tap_wave_spec} shows the tap on a PIN code digit recorded by two microphones. The spectrogram with window size 128 is shown next to the corresponding raw received signal. 

Taps can be recognised reliably with internal microphones. The sound waves propagate both in solid material i.e. the smartphone screen, and in air. Shumailov et al.~\cite{Shumailov2019HearingYT} recovered typing data using the smartphone's own microphones, finding most of the tap information in the initial burst of high-frequency waves propagating through the screen, and then using energy thresholding at relevant frequencies to find tap events. 

It is harder to locate and process taps with external microphones. Sound waves have to travel either through air, or if a table is
holding the device, through multiple solid objects. As more of the energy is
dissipated, taps have a lower signal-to-noise ratio and are harder to detect. With simple energy thresholding, a short spike can
still be observed, but the threshold has to be set low and the false positive rate will be high. It is still useful to have a set of candidate taps which can later be filtered using
more refined post-processing methods. 

Many users enable sound or vibration feedback, where the device emits a distinctive audio signal after each tap. These patterns are much easier to detect than taps. Shumailov~\cite{Shumailov2019HearingYT} explained how they can be used for tap detection, even if the delay
between the actual tap and the feedback is variable. In this work we assume a harder threat model where no feedback is emitted by the target.

\begin{figure*}
    \centering
	\includegraphics[width=0.225\textwidth]{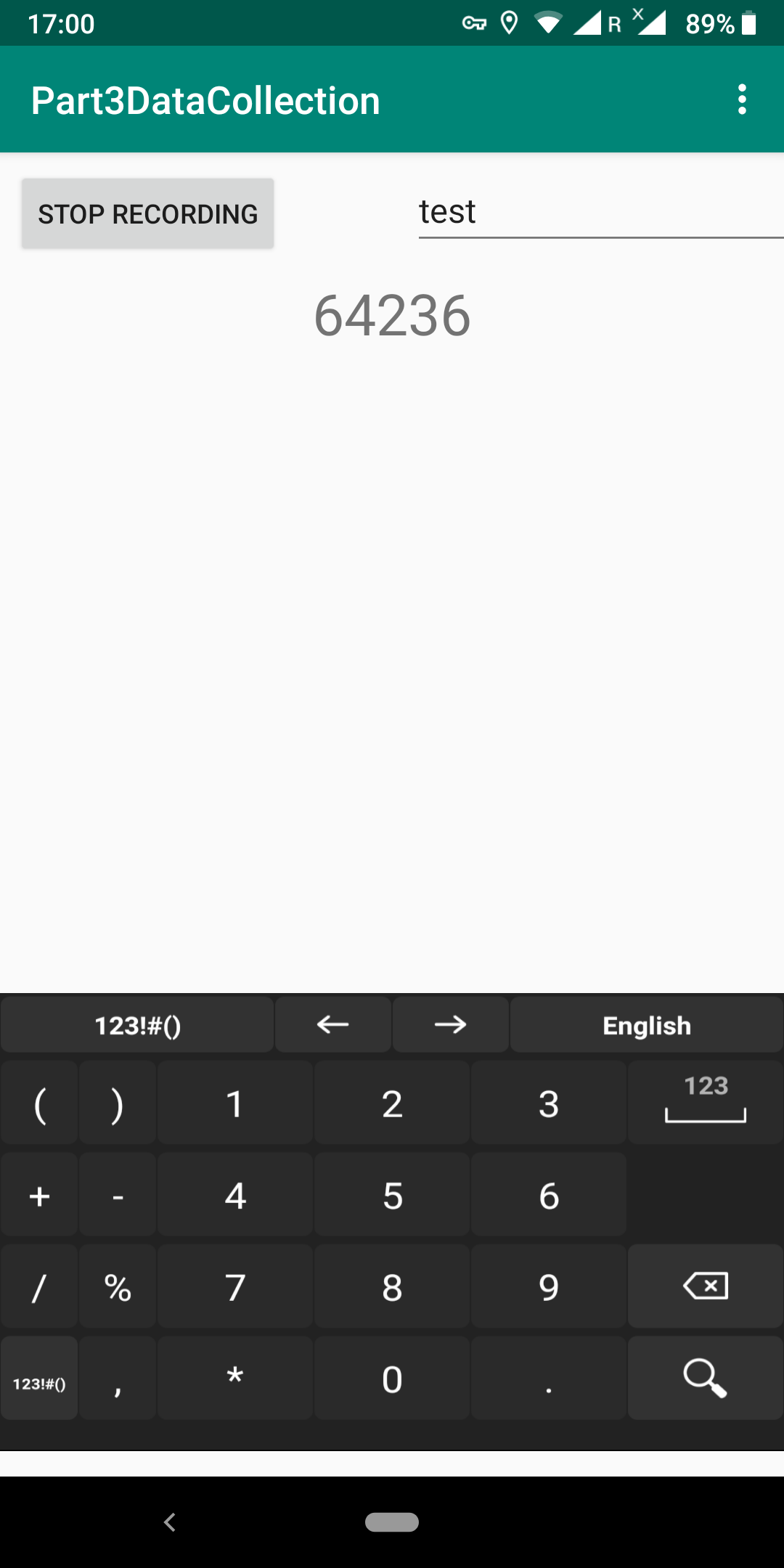}
	\includegraphics[width=0.6\textwidth]{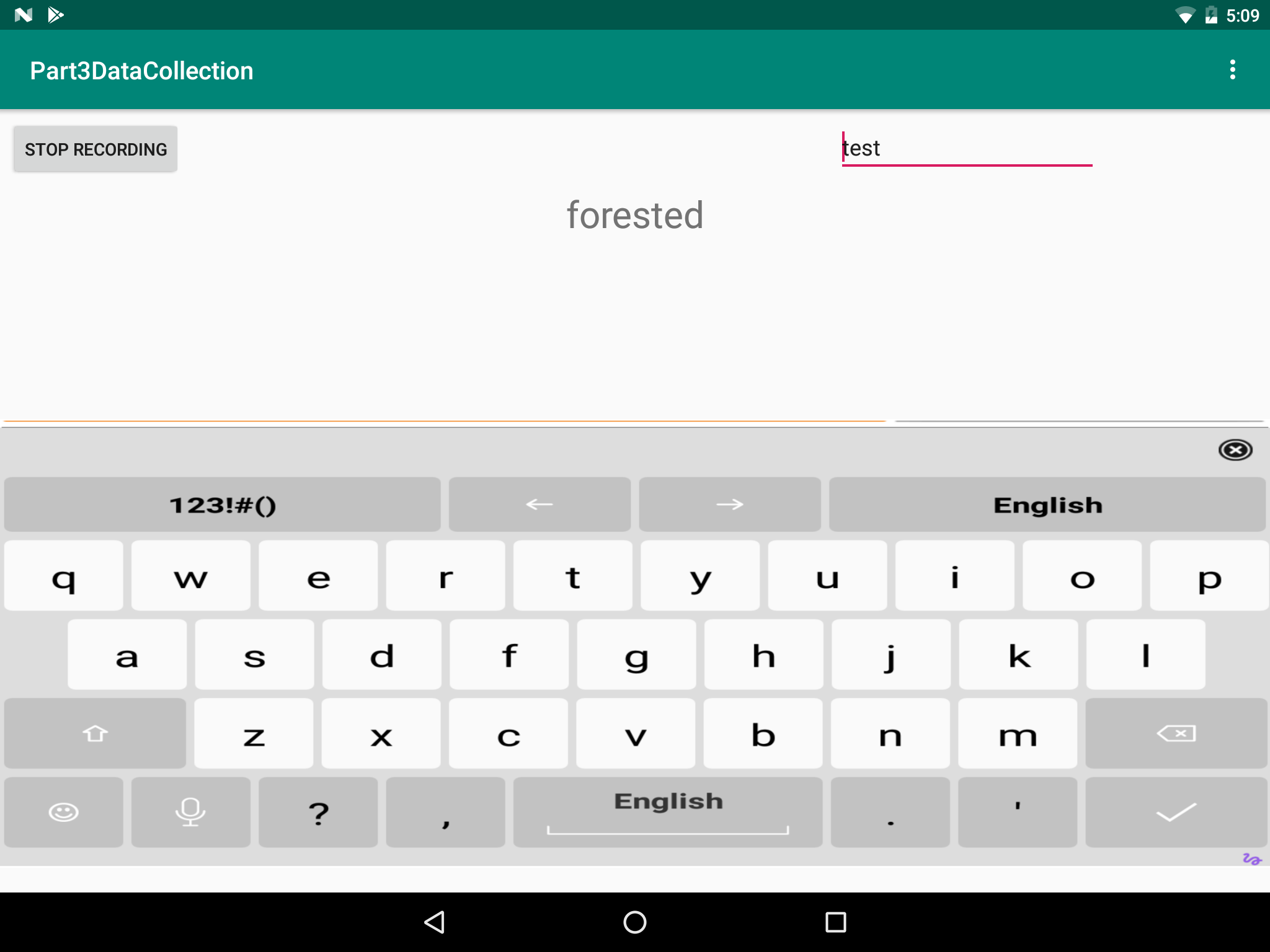}
	\caption{Application running on two devices in PIN and text entry, and in portrait and landscape
	modes}
\end{figure*}

\section{Experiment Setup}

A Raspberry Pi with a ReSpeaker 6-mic circular array was used to collect data. The Pi was running a
simple TCP server that could be told to start and stop recording and save the audio to a 6-channel
\c{.wav} file. A WiFi hotspot was set up on the Pi so that participating devices could connect to it.

The victim devices in our experiments were a HTC Nexus 9 tablet, a Nokia 5.1 smartphone and a Huawei
Mate20 Pro smartphone, all running Android 7 or above and having at least two microphones\footnote{Modern
smartphones have multiple microphones for domain-specific purposes. For example, there is
usually a microphone next to the video camera, which can be sampled
from the direction of the object being recorded. Similarly, there is usually a microphone at the
top, to denoise the microphone at the bottom that is used to collect speech during a phone call.}.
\Cref{tab:device_stats} shows some statistics for these devices. Only one device of each type was
available; we assumed that if a second identical device were used, it could be attacked using
training data from the first device. Shumailov et al. found this to be the case for Nexus 5 phones in earlier work~\cite{Shumailov2019HearingYT}. 

\begin{table*}
	\centering
	\caption{Statistics of devices used in evaluation}
	\label{tab:device_stats}
	\begin{tabular}{llcrrrr}
	\toprule
		\textbf{Vendor} & \textbf{Device Name} & ~~ & \textbf{Type} & \textbf{Size (mm)} & \textbf{Screen (mm)} & \textbf{Weight (g)}\\\midrule
		HTC&Nexus 9&&Tablet&\(228.2\times153.7\times8.0\)&\(181\times136\)&425\\
		HMD Global & Nokia 5.1&&Phone&\(151.1\times70.7\times8.3\)&\(124\times62\)&150\\
		Huawei&Mate20 Pro&&Phone&\(157.8\times72.3\times8.6\)&\(147\times68\)&189\\
		\bottomrule
	\end{tabular}
\end{table*}

The victim device was held next to the array with its screen roughly coplanar with the microphones
in the array. It was running an app that connected to the Pi and told it to start and stop
on the press of a button. At the same time, the victim device also started recording audio with 2
microphones, and an inaudible sound sequence was played to enable alignment of the two audio recordings.
The data collected from the internal microphones are not used in the attack, but just to verify the automatic labelling of the dataset, as described in \Cref{sec:autolabel},
which will be used to evaluate the attack performance. 
The victim device also recorded screen touch events with timestamps. Again, the attack does not use these data; they are only collected to verify analysis results. 

The app displayed an image of a keyboard or numeric pad, and, to ensure input
randomness, it displayed a 5-digit number or a word from an
English dictionary when the recording started. Participants were told to type the
displayed text, but not to worry about making a mistake. Sound and vibration feedback were not enabled. Audio recording was stopped and restarted after each word to reduce data loss from
overflowing buffers in recording devices. 

\subsection{The Data and Ethics}

Multiple datasets were recorded using different devices, positioning, holding techniques, and input types (PIN, text, etc). Each dataset was recorded for about half an hour, during which
250-500 taps were recorded. 

Originally, the plan was to run a user study to collect data from many participants, and the data collection
framework was designed to make this convenient and ethical. The devices in the study are provided to the participants and no applications have to be installed on the devices of the participants. 
To make sure the participants did not think they had to enter their own PIN, 5-digit PINs were solicited instead of the 4-digit PINs common in banking. The participants were advised not to talk during
the experiment, as the audio recordings might be published anonymously and recorded voice could compromise their anonymity. 

The user study got ethical approval from the university ethics review board. However due to the
COVID-19 epidemic, we were notified that it was no longer possible to recruit participants and run user studies requiring people to be in the same room and use shared equipment. Touchscreens are a possible infection transmission risk, so sharing them was not advisable during a pandemic. Thus most of the test data was provided by the researchers and their relatives, three participants in total. As a result, this paper should be considered proof-of-concept work; we show that virtual keyboard inference attacks are possible with external microphones.\footnote{We have reason to believe that the performance will only improve as more people are involved. As will be discussed later, we were careful to cross-validates the performance to make sure that there was no overfitting to data or participants.}. 

All data were recorded in a room with realistic noise levels i.e. occasional background noise from people talking and walking. We have also added additional typical background noise (podcasts playing close to the microphone array). 

\begin{figure}
	\centering
	\caption{Experiment session with the Mate20 Pro}
	\label{fig:photo}
	\includegraphics[scale=0.05,angle=90,origin=c]{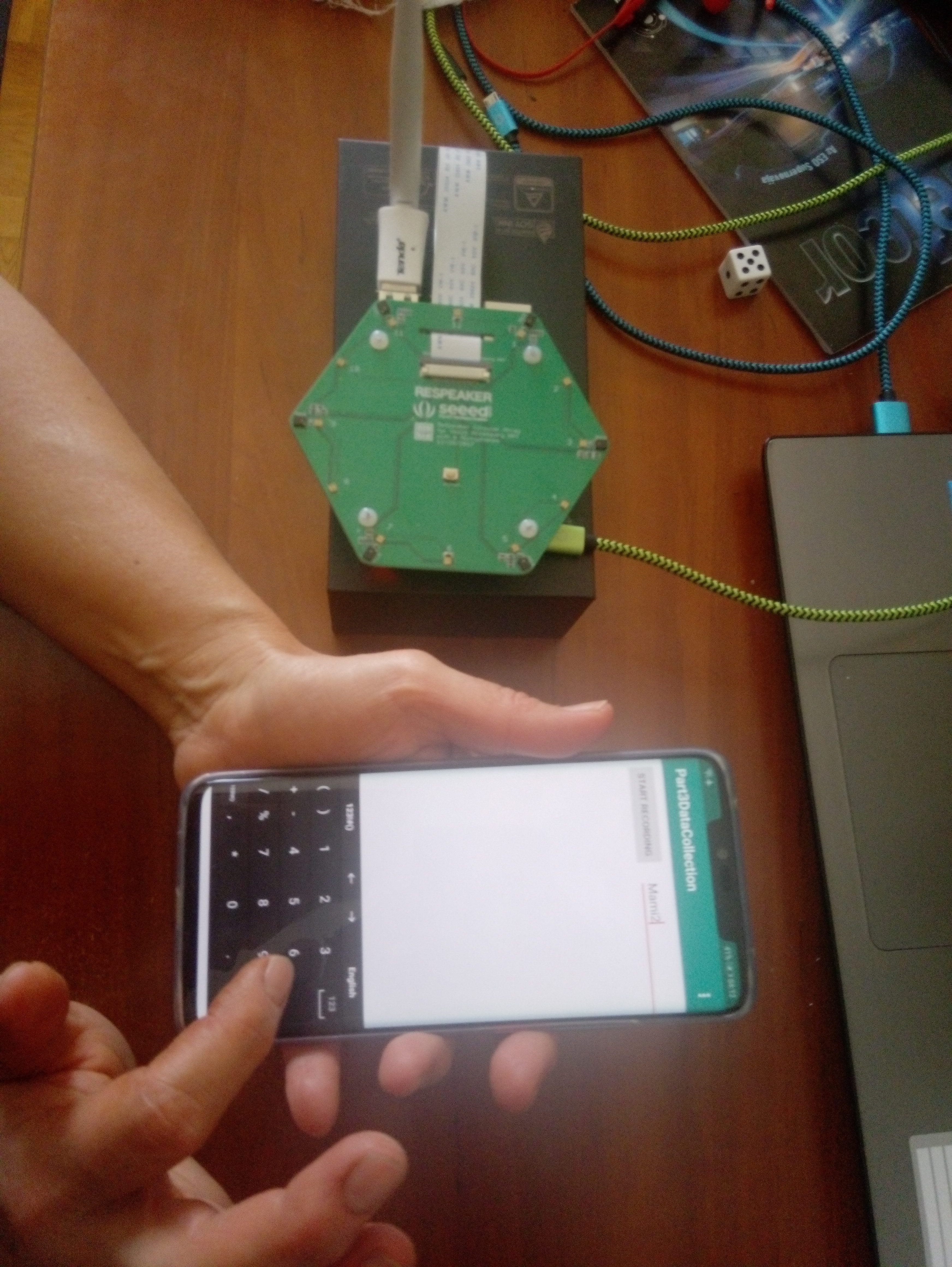}
\end{figure}

\subsection{Attack preparation and automatic labelling} \label{sec:autolabel}

In \Cref{sec:background}, we saw that a set of candidate finger taps can be recognised by looking for short wide-band spikes in the spectrogram. Shumailov et
al.~\cite{Shumailov2019HearingYT} could find most of the finger taps in internal microphone recordings
using energy thresholding. Those could then be used to build more sophisticated detection mechanisms. With external microphones, that is not possible as the false-positive rates end up dominating.
Instead we split the detection into two stages. First, we use energy thresholding to find all candidate taps; second, we use timestamping information to sift through the candidate taps. Note that the timestamping information is only used in the preparation stage of the attack to minimise the number of falsely detected taps. 

The automatic labelling can be verified by comparing the labelling of the recording on the victim
device to the labelling of the microphone array recording. The labelling on the victim device can be
trusted to be correct, since taps there have a high signal-to-noise ratio. Given a match, the external recording can be labelled manually without needing an audio recording from the victim device. 

We find that between \SIrange{40}{80}{\percent} of microphone array recordings can be labelled automatically. Manual inspection of cases where automatic labelling failed revealed three main causes. 
\begin{itemize}
    \item First, often a part of one of the recordings is missing because of an overflowing audio buffer in the recording app. In this case the recording cannot be automatically labelled because the tap events don't correspond to time offsets in the audio recording.
    \item Second, the recording on one of the devices starts too late and the synchronising sequence is missing. In this case, the recording can be labelled automatically, but its success cannot be verified.
    \item Finally in some cases automatic labelling fails because the recording is too noisy and the algorithm cannot handle this level of noise. In these cases, the labelling of the recording from the victim device can be used for the microphone array recording. 
\end{itemize} 

We find all three of the cases to be equally common in practice, making them hard to discriminate automatically. We have decided to excluded non-automatically labeled data from further evaluation. 

The automatic labelling method allows us to collect a large set of labelled data with no more manual
effort than tapping on a screen\footnote{Interestingly we find that where vibration feedback was enabled, taps can be detected without using any classifier since vibration feedback has a similar delay to the tap event.}. We have released the experiment code-base to facilitate further experimentation\footnote{\url{https://github.com/zarandya/PartIIIDataProcessing}/}.

\subsection{Tap detection} 
\label{sec:met_tapdetect}

To perform automatic labelling, the timestamps of tap events are required. The attacker does not
have access to this, so other methods are needed to detect taps in unlabelled data. To this end we train a classifier to filter the candidate taps and distinguish actual taps from false positives. 

Fourier features of tap candidates were extracted from the recordings. The features were extracted from an 512 sample segment around the taps using 128-sample long Hann windows with a 32 sample delay between the start of successive windows. 16 frequency buckets were used, with frequencies up to \SI{12000}{\hertz}. MFCC features were also extracted from
the same windows, 20 features each. 

The window size and delay were chosen because this is where the taps are most recognisable in a spectrogram by a human eye. They turned out to be the optimal shapes for tap detection.

Discriminant analysis and convolutional neural networks were tried for classification. In general, linear discriminant analysis performed better than quadratic discriminant analysis. This is somewhat surprising as it is completely unreasonable that taps and false positives have a similar covariance matrix; however LDA is reportedly flexible at bending the assumptions. LDA works best on MFCC features while CNNs work best on Fourier features. Multiple CNN configurations with varying kernel sizes and input shapes were tested. We also tried using deeper networks, but found that they didn't perform much better for the task at hand. Tap detection was evaluated using data from one, two, and all six microphones to simulate attack scenarios where the attacker has access to different numbers of microphones. We further confirmed the results with 10-fold cross validation. 

\subsection{Note on time difference of arrival}

Shumailov et al.~\cite{Shumailov2019HearingYT} relied heavily on time difference of arrival (TDoA) for their performance. We have evaluated a large number of different approaches for TDoA but were not able to find any that gave useful distance estimates, regardless of what source-localisation techniques we tried. We did however find that points from different presses of the same key were placed near a line that passed through the centre of the
array and went in the direction of the key. So direction estimation may still be usable, but precision requirements make it hard to use for virtual keyboards. We hypothesise that waves passing through materials allowed Shumailov et al. to get more precise measurements. 

\subsection{Tap classification}

Once a segment of audio is decided to be a tap, we want to determine which key on the keyboard was pressed. In the rest of this dissertation, ``classification'' refers to assigning keys to taps
(or tap candidates, assuming they are taps), using a CNN or LDA classifier. ``Classifier'' may refer
to a classifier used either for detection or for classification, depending on context. 

As with tap detection in~\Cref{sec:met_tapdetect}, classification was done using linear
discriminant analysis on MFCC features and neural networks on Fourier features. Instead of two
classes, there was now one for each key. Shumailov et al.~\cite{Shumailov2019HearingYT} found that
LDA with Cepstrum features performed best for internal microphones. 

Again, the most relevant information was contained in a 512 sample segment near the start of the
tap, which has the spike of high frequencies. Unlike in the previous work, smaller MFCC windows did not
perform well here, so a \(1024\)-sample window was used. 
%

It should be mentioned that we expect lower accuracy, because we have more classes and fewer
training data points. However, our data still indicate that it is possible to perform the attack. Here, we report the number of taps classified correctly on the first, second and third tries.

\section{Evaluation} 

\subsection{Signal strength}

\begin{figure}[h]
	\centering
	\scalebox{0.55}{%
		\input{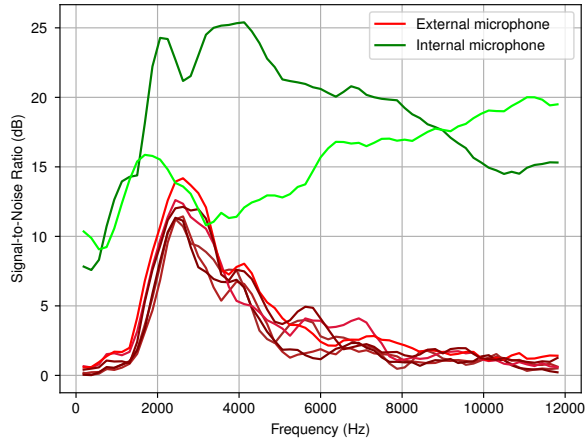}
	}
	\caption{Average Signal-to-Noise Ratio (in decibels) of a tap on a Nokia 5.1 when recorded on the phone
	(green) and the microphone array (red) at each frequency. Different lines are different
microphones.}
	\label{fig:snr}
\end{figure}

The Signal-to-Noise Ratio of a tap recorded on the internal microphone of a victim device is around
\SIrange{15}{20}{\deci\bel}. On the external array, the volume is much lower, but the taps and noise
do not have the same frequency distribution. As a result, the Signal-to-Noise ratio varies by
frequency. As seen on~\Cref{fig:snr}, the SNR peaks between
\SIrange{2000}{3500}{\hertz} around \SI{10}{\deci\bel} and is still significant in the
\SIrange{3500}{4500}{\hertz} range. Under these ranges, the noise is too high, while above them
the signal is too low. 

\begin{figure}
	\centering
	\scalebox{0.55}{%
		\input{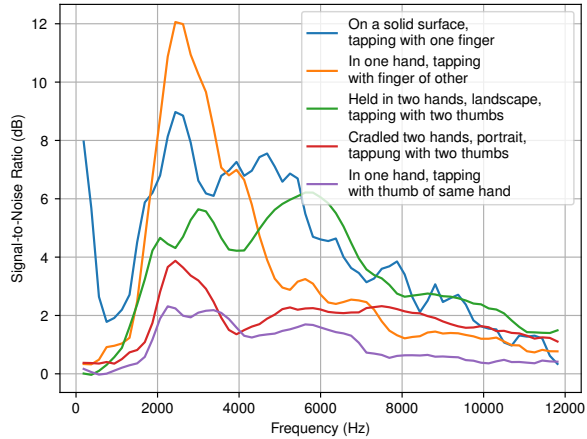}
	}
	\caption{Dependence of Signal-to-Noise ratio on different ways of holding the Nokia 5.1 device}
	\label{fig:snr_holding}
\end{figure}

Both signal strength and its frequency distribution depend on which device is used, on the way it is held, and on the way it is tapped. As seen on~\Cref{fig:snr_holding}, it is better if the device is
held on a table or in a different hand than the one doing the tapping, as this gives harder impact. Fortunately,
these are the most common ways of holding modern touch-screen devices. When a device is held in two hands, the signal
is better in landscape mode than portrait mode, but it peaks at higher frequencies. When held in one
hand and tapped with the same hand -- as used to be common for older, smaller devices -- the signal is
weak, automatic labelling fails on most data, and classification is only slightly better than random guessing. 

\subsection{Tap detection}

The precision and recall of the best performing LDA and CNN classifiers for each dataset are shown
in~\Cref{tab:tap_detect}. Recall is the proportion of positives found, precision is the proportion
of true positives among those marked positive. 

\begin{figure}
	\centering
	\scalebox{0.7}{%
		\subimport{figs/}{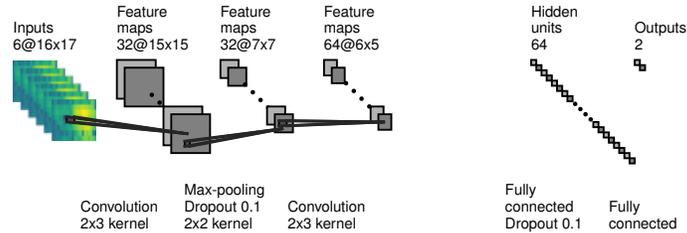}
	}
	\caption{Best performing convolutional neural network for tap detection}
	\label{fig:cnn}
\end{figure}

%


With both LDA and CNNs, tap recognition performs above \SI{75}{\percent} in precision and above
\SI{80}{\percent} in recall. CNNs slightly outperform LDA in both metrics.  

When only two microphones are available (in a setup similar to Google Home), recall drops to the
\SIrange{70}{75}{\percent} range when using LDA. CNNs perform much worse with only two microphones,
the recall drops to around \SIrange{65}{70}{\percent}. Precision consistently stays above
\SI{80}{\percent}.

With only one microphone, recall is around \SI{65}{\percent} for LDA and \SI{55}{\percent} for CNNs. 

\subsection{Classification of numeric digits} \label{sec:eval_classify}

\Cref{tab:tap_classify} shows the proportion of taps correctly classified on the first, second
and third guess for each dataset.

\begin{table*}[t]
    \centering
    \adjustbox{max width=\textwidth}{
    \begin{tabular}{lllllcrr}
        \toprule
		Device&holding&distance&orientation&keyboard and entry&method&precision&recall\\\midrule
		Nexus 9 Tablet&in one hand&\SI{20}{\centi\metre}&portrait&5-digit PINs&MFCC+LDA&\SI{83}{\percent}&\SI{76}{\percent}\\ 
		\vspace{2mm}
		Nexus 9 Tablet&in one hand&\SI{20}{\centi\metre}&portrait&5-digit
		PINs&Fourier+CNN&\SI{85}{\percent}&\SI{77}{\percent}\\
		Nokia 5.1 Phone&in one hand&\SI{15}{\centi\metre}&portrait&5-digit
		PINs&MFCC+LDA&\SI{81}{\percent}&\SI{76}{\percent}\\ 
		\vspace{2mm}
		Nokia 5.1 Phone&in one hand&\SI{15}{\centi\metre}&portrait&5-digit
		PINs&Fourier+CNN&\SI{87}{\percent}&\SI{80}{\percent}\\
		Mate20 Pro Phone&in one hand&\SI{20}{\centi\metre}&portrait&5-digit
		PINs&MFCC+LDA&\SI{87}{\percent}&\SI{77}{\percent}\\
		Mate20 Pro Phone&in one hand&\SI{20}{\centi\metre}&portrait&5-digit
		PINs&Fourier+CNN&\SI{91}{\percent}&\SI{82}{\percent}\\
		\midrule
		Nexus 5 Tablet&on a table&\SI{20}{\centi\metre}&portrait&English
		words&MFCC+LDA&\SI{93}{\percent}&\SI{77}{\percent}\\
		Nexus 5 Tablet&on a table&\SI{20}{\centi\metre}&portrait&English
		words&Fourier+CNN&\SI{91}{\percent}&\SI{78}{\percent}\\\bottomrule
	\end{tabular}
	}
	\caption{Tap detection experiments} \label{tab:tap_detect}
\end{table*}

\begin{table*}[t]
    \centering
    \adjustbox{max width=\textwidth}{
    	\begin{tabular}{lllllcrrr}
    	\toprule
		Device&holding&distance&orientation&keyboard and entry&method&accuracy&2nd guess&3rd guess\\\midrule
		Nexus 9 Tablet&in one hand&\SI{20}{\centi\metre}&portrait&5-digit
		PINs&MFCC+LDA&\SI{41}{\percent}&\SI{61}{\percent}&\SI{73}{\percent}\\ 
		\vspace{2mm}
		Nexus 9 Tablet&in one hand&\SI{20}{\centi\metre}&portrait&5-digit
		PINs&Fourier+CNN&\SI{31}{\percent}&\SI{48}{\percent}&\SI{63}{\percent}\\
		Nokia 5.1 Phone&in one hand&\SI{15}{\centi\metre}&portrait&5-digit
		PINs&MFCC+LDA&\SI{48}{\percent}&\SI{66}{\percent}&\SI{76}{\percent}\\
		\vspace{2mm}
		Nokia 5.1 Phone&in one hand&\SI{15}{\centi\metre}&portrait&5-digit
		PINs&Fourier+CNN&\SI{37}{\percent}&\SI{57}{\percent}&\SI{69}{\percent}\\
		Mate20 Pro Phone&in one hand&\SI{20}{\centi\metre}&portrait&5-digit
		PINs&MFCC+LDA&\SI{37}{\percent}&\SI{54}{\percent}&\SI{66}{\percent}\\ 
		Mate20 Pro Phone&in one hand&\SI{20}{\centi\metre}&portrait&5-digit
		PINs&Fourier+CNN&\SI{28}{\percent}&\SI{52}{\percent}&\SI{66}{\percent}\\
		\midrule
		Nexus 9 Tablet&on table&\SI{20}{\centi\metre}&portrait&English
		words&MFCC+LDA&\SI{47}{\percent}&\SI{61}{\percent}&\SI{68}{\percent}\\
		\vspace{2mm}
		Nexus 9 Tablet&on table&\SI{20}{\centi\metre}&portrait&English
		words&Fourier+CNN&\SI{34}{\percent}&\SI{51}{\percent}&\SI{60}{\percent}\\
		Nexus 9 Tablet&on table&\SI{15}{\centi\metre}&landscape&English
		words&MFCC+LDA&\SI{41}{\percent}&\SI{51}{\percent}&\SI{59}{\percent}\\
		\vspace{2mm}
		Nexus 9 Tablet&on table&\SI{15}{\centi\metre}&landscape&English
		words&Fourier+CNN&\SI{32}{\percent}&\SI{46}{\percent}&\SI{54}{\percent}\\
		Nokia 5.1 Phone&in two hands&\SI{15}{\centi\metre}&landscape&English
		words&MFCC+LDA&\SI{42}{\percent}&\SI{55}{\percent}&\SI{61}{\percent}\\
		Nokia 5.1 Phone&in two hands&\SI{15}{\centi\metre}&landscape&English
		words&Fourier+CNN&\SI{29}{\percent}&\SI{43}{\percent}&\SI{51}{\percent}\\\bottomrule
	\end{tabular}
	}
    \caption{Tap classification experiments. Accuracy is the proportion of taps correctly classified on first guess} \label{tab:tap_classify}
\end{table*}



The classification accuracy is the proportion of taps that are correctly classified on the first
guess. Taps correctly classified on the second or third guesses are also of interest. 
The accuracy is variable between datasets, but is around or above \SI{30}{\percent} for CNNs and
\SI{40}{\percent} for LDA. In comparison, Shumailov et al.~\cite{Shumailov2019HearingYT} could
correctly classify around \SI{64}{\percent} of digits on the first try using internal
microphones\footnote{They also only had 9 classes instead of 10.}. 

\begin{figure}[t]
	\centering
	\scalebox{0.55}{%
		\subimport{figs/}{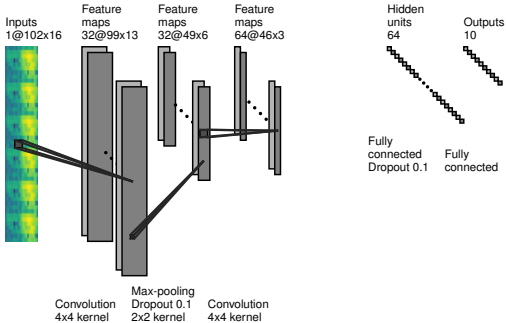}
	}
	\caption{Best performing convolutional neural network for classification}
	\label{fig:cnn_classify}
\end{figure}

Classification can also be done with data from two microphones, with slightly worse results. Using
LDA, the accuracy is between \SIrange{35}{40}{\percent} for each dataset, and the second and third
guesses are also very close to the 6 mic result. With one microphone, the accuracy will be around
\SI{32}{\percent}. With CNNs, the accuracy is between \SIrange{25}{30}{\percent} for two
microphones, again close to the 6 microphone results. With one microphone, it performs worse, around
\SI{15}{\percent}.

\subsection{Classification of letters}

Letters can be classified in the same way as numeric digits. The same LDA and CNN classifiers work
well for them, with the same features and configurations, except for the number of output classes
(and hence the final CNN layer).

The test data contains entries of English words. The taps are distributed according to the frequency
of letters in English, but are classified individually, so as far as the classifier is concerned
the distribution of a tap is independent of the previous letter. A language model can be
used to improve classification. 

\begin{figure}
	\centering
	\scalebox{0.55}{%
	\subimport{figs/}{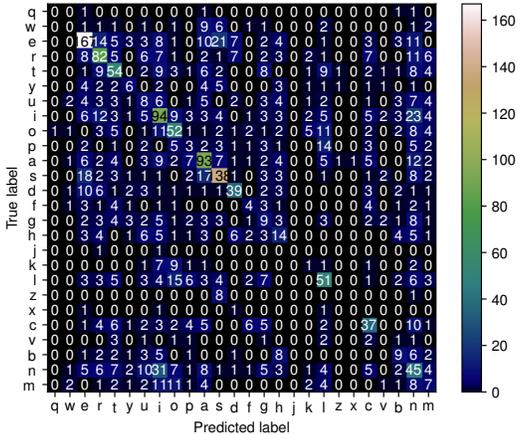}
	}
	\caption{Confusion Matrix of classified key presses on the Nokia 5.1 held in two hands in
	landscape mode.}
	\label{fig:conf_matrix}
\end{figure}

Letters are often misclassified as other letters adjacent on the keyboard
(see~\Cref{fig:conf_matrix}). This is similar to the common problem with smartphones that the keys
are too small and users often tap the wrong one. Autocorrect keyboards are used to fix this problem
using a dictionary; an attack based on a dictionary is presented in~\Cref{sec:guess_word}. 

The two most common letters in our test dataset, \c e and \c s each amount for \SI{10}{\percent} of all characters\footnote{Our test words were drawn from the Unix
\c{words} file which includes every word once, and we ignored word frequency. The file included the plural and third person form of words, so the letter \c s was
overrepresented.}, so the random guesser with the
best accuracy would only choose these letters. 
The classifier does reasonably well at separating \c e and \c s, even though they are adjacent on
the keyboard. It does not do well at identifying rate characters which are next to a common
character. For example, a \c w is almost always classified as an \c s or an \c a. About \SI{30}{\percent} of taps can be correctly classified using CNNs, and \SI{40}{\percent} using
LDA. 

Phones are often held in two hands when typing text. In this case, classification of letters only
works well in landscape mode. In portrait mode, the keys are too close to each other, and cradling
the device in two hands has a low signal strength. 

\section{Direction-of-arrival estimation}

\begin{figure}[h]
	\centering
	\scalebox{0.55}{%
		\input{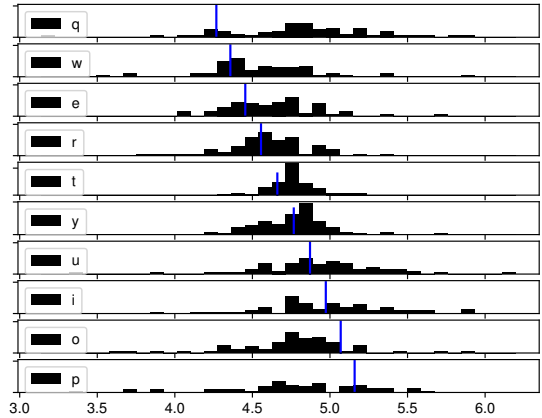}
	}
	\caption{Histogram of predicted directions (in radians) using TDoA measurements for taps on
	characters in the top row of the keyboard (\c{qwerty\ldots}). The blue line shows the true direction towards the middle of the key.}
	\label{fig:tdoa_hist}
\end{figure}

\Cref{fig:tdoa_hist} shows histograms of estimated directions of arrival for each letter. 
The most likely direction bin bears a close resemblance to the direction of the character. The
variance is large though, and the classes cannot be separated. Therefore this
information is not sufficient to classify the taps, even in one row of the keyboard. It still
provides a probability distribution for each key, which may be a useful additional feature in 
classification. 

In an experiment performed on recordings of words typed in landscape mode, the estimated direction of
arrival was added as an additional feature to MFCC features and classified using LDA. This slightly
improves the classification accuracy, but the difference is not significant. Taps misclassified by
LDA were already mostly classified as adjacent keys, and an inaccurate direction of arrival
estimation does not add much to that. Similar results can be achieved by adding the time differences
of arrival between microphone pairs as LDA features. 

In these experiments, the victim device was held still. If an attacker cannot exactly reproduce the
device positioning of the victim when recording test data, DoA will not be a useful feature for
classification.

\subsection{Effect of distance}

\begin{figure}
	\centering
	\scalebox{0.55}{%
		\input{figs/snr-tdist.5.pgf}
	}
	\caption{Average Signal-to-Noise ratio of taps on the Nexus 9 Tablet recorded on the microphone
	array at different distances. (Different lines with similar colours are different microphones in
	the array at that distance)}
	\label{fig:snr_tdist}

	\scalebox{0.55}{%
		\input{figs/tdist-results.pgf}
	}
	\caption{Effect of distance on recall (top) and classification accuracy of numeric digits
	(bottom). Exp.~1 is trained with data from the same distance. Exp.~2 is trained with data from
	different distances.}
	\label{fig:tdist_result}
\end{figure}

The Signal-to-Noise ratio decreases with distance, as can be seen on \Cref{fig:snr_tdist}. 
\SI{55}{\centi\metre} away from the array, it peaks at \SI{2}{\deci\bel}. At this point, the
automatic labelling only succeeds on \SI{20}{\percent} of samples. This is
not sufficient to evaluate detection and classification, since much more data will be needed for
training the classifiers; it furthermore introduces a bias towards the few data points where signal
volume is decent despite the distance. 

Instead, we use all data, and where the labelling based on the external microphone fails, fall back
to the labelling from the recording on the internal microphone. This means there are incorrectly
labelled recordings in the training set, due to the other two causes of automatic labelling failure
described in \Cref{sec:autolabel}. Since those are caused by recording errors, the proportion of
them in the entire dataset can be assumed independent of the distance and results at different
distances can be compared. 

In one experiment, tap detection and classification were performed independently for each distance.
In a second experiment, classifiers were trained using data for all distances, and only test data was
split by distance. As can be seen on \Cref{fig:tdist_result}, both recall and classification accuracy drop
rapidly, at \SI{30}{\centi\metre}, under \SI{60}{\percent} of taps will be detected, which makes the
attack hardly feasible at that distance. The second experiment performs better due to more training
data, this suggests that it is not necessary to know the position of the victim device exactly for
gathering data. 

\subsection{PIN guessing}

In the previous sections, we have seen that both LDA and neural networks can provide reasonable
guesses for individual taps. However, to reconstruct a whole PIN, all taps have to be detected
and classified correctly. With a 5-digit PIN, this is almost never the case.

Often an attacker can only make a limited number of guesses. Studies of side-channel attacks therefors typically report what proportion of inputs can be reconstructed in a fixed number of guesses. This enables a security engineer how far the entropy of the keyspace is eroded by the side channel under study. In this section,
5-digit PINs are reconstructed from candidate taps with classifiers performing detection and
classification. The attacker will consider each option as a guess, assign them a likelihood, and
perform guessing in order of likelihood.

The LDA classifiers not only return the most likely class for a tap, they also give a probability
distribution of classes. The probability is derived from Bayes' theorem and the LDA assumptions. For
CNNs, a probability distribution can be obtained from the softmax function on the final layer. The
likelihood of a guess can be computed by multiplying the probabilities obtained from the classifier
for each digit.

Tap detection does not always succeed, but detection classifiers also give probabilities for each
candidate that it is a tap. The classification step gives conditional probabilities for each class
given that it is a tap. To consider multiple candidates, the probability that each
candidate tap is a tap is also multiplied by the likelihood of that guess. The
probability that all other candidates are false positives must also be considered. 

The likelihood of a guess is therefore the product of the the probability
that each candidate tap is indeed a tap, the probability that everything else is not a
tap, and the probability that each tap is a given digit. In practice, logarithmic probabilities
are used for numerical stability. 

Then the number of inputs at least as likely as the real input is
taken as the number of guesses. Note that the guesses include not just the PIN but exactly when the
keys were pressed, as multiple guesses may correspond to the same PIN. 

The computational complexity of computing every guess is high. However candidates can be sorted so that low probability guesses are tested later\footnote{Candidates have to be
	sorted in decreasing order by \(\frac{p_n\max_ic_{n,i}}{1-p_n}\) where \(p_n\) is the probability
	that it is a tap, and \(c_{n,i}\) is the probability that it is the class \(i\) assigned by
	the classifier. This is because they contribute \(p_n\max_ic_{n,i}\) to the likelihood of the
	guess if they are chosen, and \(1-p_n\) if
another candidate is chosen instead.}. If choosing a candidate
results in no guesses more likely than the truth, later candidates do not have to be
tried. So only guesses with higher likelihood
are computed (and a few extra). If this number is still too high, then the algorithm exits prematurely and the side channel analysis is ineffective at improving the guessing probability.

\begin{figure}
	\centering
	\scalebox{0.55}{%
		\input{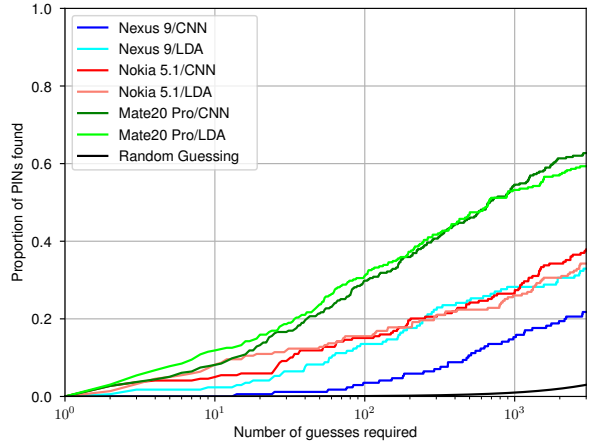}
	}
	\caption{Proportion of 5-digit PINs classified using LDA and CNN with a limited number of
	guesses. (Note the logarithmic axis) }
	\label{fig:guess}
\end{figure}

\Cref{fig:guess} shows effectiveness of guessing 5-digit PIN numbers. Many PINs can be found in a small number of guesses, but later the curve flattens and over half of possible PINs cannot be found in 3000 guesses.

\begin{figure}
	\centering
	\scalebox{0.55}{%
		\input{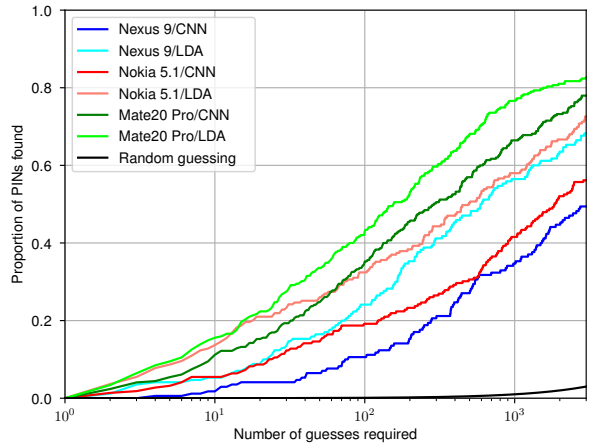}
	}
	\caption{Proportion of 5-digit PINs classified using LDA and CNN with a limited number of
	guesses, with the additional assumption that taps can be detected correctly. (Note the
	logarithmic axis)}
	\label{fig:guess_skipdetect}
\end{figure}

In almost all cases when the PIN is found, tap detection correctly finds all taps, and in the remaining cases, all taps are among the 6-8 most likely candidates. On the other hand, if not all taps
were correctly detected, and the ones that were detected did not have a much higher likelihood than false
positives, the PIN could not be guessed. Since recall is in the
\SIrange{75}{80}{\percent} range, in many recordings not all taps will be found correctly. 

Thus guessing performs better if taps can be detected correctly, and do not have to be detected by
classifiers. Here, false positive candidates are ignored, and only the probabilities of each digit
are used to compute the likelihood. \Cref{fig:guess_skipdetect} shows the required number of guesses
with this assumption\footnote{It is worth noting that in practice this attack will perform much better, because humans pick a very limited set of PIN-codes and passwords~\cite{bonneau2012thescience,wang2017understanding}.}.

A similar guessing attack was performed on internal microphones by Shumailov et al.~\cite{Shumailov2019HearingYT}.   \Cref{tab:guess_compare} compares the results. 

\begin{table*}
	\centering
	\caption{Comparison of guessing attack with internal microphone attack by Shumailov et al.~\cite{Shumailov2019HearingYT}. 4-digit PINs are used in both attacks, but our attack uses ten digits while they use nine.}
	\label{tab:guess_compare}
	\begin{tabular}{lrrrrrrlclc}
	    \toprule
		&\multicolumn{2}{c}{Nexus 5}&\multicolumn{2}{c}{Nokia 5.1}&\multicolumn{2}{c}{Mate20
		Pro}&&Nexus 5~\cite{Shumailov2019HearingYT} && Simon and Anderson~\cite{simonanderson2013pinskimmer}\\
		&CNN&LDA&CNN&LDA&CNN&LDA&~&internal mics && mic+camera\\\midrule
		10
		guesses&\SI{4}{\percent}&\SI{14}{\percent}&\SI{15}{\percent}&\SI{25}{\percent}&\SI{19}{\percent}&\SI{26}{\percent}&&\SI{72}{\percent} && \SI{61}{\percent}\\
		20
		guesses&\SI{9}{\percent}&\SI{25}{\percent}&\SI{23}{\percent}&\SI{28}{\percent}&\SI{31}{\percent}&\SI{34}{\percent}&&\SI{81}{\percent}&& \SI{84}{\percent}\\
		\bottomrule
	\end{tabular}
\end{table*}

\subsection{Dictionary guessing} 
\label{sec:guess_word}

Text can also be guessed in a similar way, but the number of guesses required will be higher due to
there being more characters, variable string length, and other factors. If the entry is drawn from a
language, knowledge of the language statistics can be used to improve guessing. 

In this attack, a dictionary is used to find the word. For each word in the dictionary, likelihood
(as described in the previous section) is computed, and the number of words with a better likelihood
than the actual word is counted. 

\begin{figure}
	\centering
	\scalebox{0.55}{%
		\input{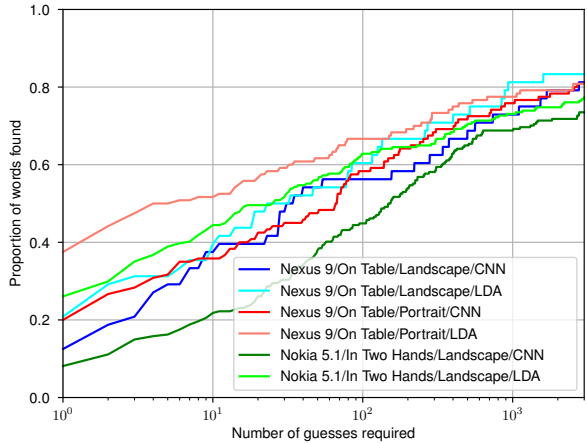}
	}
	\caption{Proportion of dictionary words found using LDA and CNN with a limited number of
	guesses, with the additional assumption that taps can be detected correctly. (Note the
	logarithmic axis)}
	\label{fig:guess_word}
\end{figure}

The attack only works well if all taps are correctly detected, which is almost never the case for
non-short words. Thus instead, the attack was evaluated with skipping the detection step and
assuming it was successful, and only performing classification on actual taps. With this assumption, the attack
works well, many words can be reconstructed in a low number of guesses as can be seen
on~\Cref{fig:guess_word}. This shows that classification works well even when
detection does not, and the attack would benefit from an improved method for detection.

While users were asked to type dictionary words, they often made mistakes. In this attack, their
typos were corrected to the most similar dictionary word. The attack could find mistyped words
where the wrong key was pressed, but could not do that in the presence of missing keystrokes. 

\subsection{Attacking other users}

In the previous sections, the training data always contained data from the same users as the test
data. Due to the pandemic, data could be collected from only a limited number of different users
although in a variety of conditions (device, holding, distance, entry). 

On a set of PINs entered on the Mate20 Pro by one user, CNN and LDA classifiers were trained, which
were used to attack PINs entered on the same device by another used. 

In this experiment, \SI{35}{\percent} of taps could
be classified correctly using MFCC and LDA on 5 microphones (the sixth microphone failed before the
experiment). This is slightly worse than the cross-validation results presented
in~\Cref{sec:eval_classify}, but still good. The second and third guesses also performed well,
finding the digit at \SI{53}{\percent} and \SI{66}{\percent} of taps, which is even closer to
cross-validation results. The CNN classifier, performed poorer on this dataset, finding the digit
only \SI{22}{\percent} of times. 

Tap detection performed as well as with cross-validation using both
CNN as LDA, and results using fewer microphones were also similar.

\section{Conclusions} 

In this paper, we have shown that inference attacks against virtual keyboards can work with external microphones. While previous work which showed that the microphones in a device can be used to deduce PINs and passwords entered on its screen, we have now shown that a smart speaker such as Alexa could in theory snoop on PINs or text entered on a nearby phone or tablet. 
Up to around \SI{40}{\percent} of taps can be correctly classified on the first guess
for both numerical and text keyboards. Linear Discriminant Analysis with MFCC features works best
for this, while Convolutional Neural Networks with Fourier features are also suitable. The effect would be to greatly reduce the number of PINs or passwords that had to be tried in order to unlock a device. This, we argue, is a practical way to measure the information leakage through such a side channel.

The weak point is the problem of tap detection. While a high proportion -- about
\SIrange{75}{80}{\percent} -- of taps can be detected, any increase in the number of false
positives means that the number of PIN guesses increases exponentially. With internal microphones, taps could be
detected using energy thresholding~\cite{Shumailov2019HearingYT}. With external microphones it is not possible to
add this kind of artificial mitigation, but the environment provides the false positives. 

When snooping on text entry, around \SI{40}{\percent} of English words can be detected using an acoustic dictionary attack. However, this
attack struggles too without correct tap detection. For our relatively small dictionary, in the
worst case scenario it would need about ten thousand guesses to find a majority of common words. The
complexity grows when larger dictionaries are assumed and new heuristics become necessary. 

To mount a more practical attack, the adversary would need a more accurate way to detect taps, such
as one based on vibration feedback (though this can be countered if vibration feedback is
artificially randomised). If detection is successful, taps can be classified, even taps
that are hard to detect otherwise. 

\section{Countermeasures}

While the attack is still hard to perform due to imperfect detection, mitigations must be discussed. 

In their seminal paper on acoustic side-channel attacks on mechanical keyboards, Asonov and Agrawal~\cite{Asonov2004KeyboardAE} proposed
using a more silent keyboard, such as a touchscreen. In this project, we have seen that touchscreens
are also susceptible to the attack, and not just by microphones physically mounted on the same device as the target touchscreen. 

Since the principal problem is tap detection, mobile vendors could try injecting false positives into
the data stream by playing silent tap-like sounds randomly while the keyboard is open, in a way that
users can not hear. A significant number of false positives will make the guessing attack infeasible. 


The attack relies on the attacker having access to an identical device. Many users
use phone cases or screen protectors to protect their phone from mechanical damage; these can also
alter tap acoustics and may provide some measure of protection against acoustic side-channel snooping. 

Regarding smart speakers, this shows that Amazon and Google were prudent in not allowing third-party skills to access access raw audio recordings. However, Apple and Google are not the only firms selling consumer electronic devices that contain MEMS microphones and support third-party apps that may be untrustworthy.

\section{Future directions}

This work is the first to demonstrate an acoustic attack on touchscreens using external microphones. It is still a proof-of-concept attack; its significance is to show that the possibility of a side channel that previous researchers had disregarded or even dismissed. 

The largest problem is discriminating between actual taps on the target screen and other quiet sounds in the room. More sophisticated acoustic approaches to
tap detection might improve significantly on our results. In particular applications, other
methods that do not rely on the intrinsic acoustic signatures of taps must be considered, such as
vibration feedback or even components that are not purely acoustic. For example, a nearby software radio might be used to listen for Tempest signals generated by the target device when processing tap inputs. 

Our work was hindered by the coronavirus pandemic which limited the number of test users. Cross-user attacks must be verified with more users at some later time, to see whether the attack can be improved by a training set
containing data from more users. Hopefully the classifiers could learn more
device-specific and user-agnostic filters, which may be more useful for attacking a user not in the
training set. One might also discover classifiers that work across devices. We might also improve the attack by just collecting more data; in our experiments,  the attack kept improving as more data was
added. 

Countermeasures should also be tested. One could try to emit false positives from a microphone, test
whether the classifiers can tell them apart from real taps, and evolve better false positive taps. The distorting effects of phone cases
should also be tested, and whether the attacker can overcome them by training with multiple cases.

\clearpage

{\small
\bibliographystyle{abbrv}
\bibliography{refs}
}

\end{document}